\pdfoutput=1
\documentclass[pdftex]{pasj00}
\begin{document}
\SetRunningHead{M. Sawada et al.}{X-ray observations of Sgr D}
\Received{2008/03/31}%{yyyy/mm/dd}
\Accepted{}%{yyyy/mm/dd}

\title{X-Ray Observations of the Sagittarius D H \emissiontype{II} Region 
\\ toward the Galactic Center with Suzaku}
%%% begin:list of authors
% Do NOT capitalize all letters in "textsc".
\author{Makoto~\textsc{Sawada},\altaffilmark{1} 
Masahiro~\textsc{Tsujimoto},\altaffilmark{2}\thanks{Chandra Fellow} 
Katsuji~\textsc{Koyama},\altaffilmark{1} 
Casey~J.~\textsc{Law},\altaffilmark{3}\\ 
Takeshi~Go~\textsc{Tsuru},\altaffilmark{1} 
and Yoshiaki~\textsc{Hyodo}\altaffilmark{1}}
  %\thanks{}}
\affil{$^1$Department of Physics, Graduate School of Science, 
Kyoto University \\
Kitashirakawa Oiwake-cho, Sakyo-ku, Kyoto 606-8502}
\affil{$^2$Department of Astronomy \& Astrophysics, 
Pennsylvania State University\\ 525 Davey Laboratory, 
University Park, PA 16802, USA}
\affil{$^3$Astronomical Institute ``Anton Pannekoek'', University of Amsterdam\\
Kruislaan 403, 1098 SJ Amsterdam, The Netherlands}
\email{sawada@cr.scphys.kyoto-u.ac.jp}

%% `\KeyWords{}' always has to be placed before `\maketitle'.
 %Do NOT move this preamble from here!

\maketitle

\begin{abstract}
%Please read ``IMPORTANT NOTICE'' carefully before preparing a manuscript. 
We present a Suzaku X-ray study of the Sagittarius D (Sgr\,D) H\emissiontype{II} 
region in the Galactic center region. Two 18\arcmin$\times$18\arcmin\
images by the X-ray Imaging Spectrometer (XIS) encompass the entire Sgr\,D complex. 
Thanks to the low background, XIS discovered two diffuse sources 
with low surface brightness and 
obtained their high signal-to-noise ratio spectra. 
One is associated with the core of the Sgr\,D H\emissiontype{II} region, arising from the 
young stellar cluster. The other is a new object in the vicinity of the region. 
We also present 3.5~cm and 6.0~cm radio continuum maps of the new source 
using the 100~m Green Bank Telescope. We conclude that the source is 
a new supernova remnant (SNR; G1.2--0.0) based on: (1) the 0.9$\pm$0.2~keV 
thermal X-ray spectrum with emission lines from highly ionized atoms; 
(2) the diffuse nature with an apparent extent of $\sim$10~pc 
at the Galactic center distance inferred from the X-ray absorption 
($\sim$8.5$\times$10$^{22}$~cm$^{-2}$); and 
(3) the nonthermal radio continuum spectral index ($\sim -$0.5).  
Our discovery of an SNR in the Sgr\,D H\emissiontype{II} region leads to 
a revision of the view of this system, which had been considered to be a 
thermal H\emissiontype{II} region and its environment.
\end{abstract}

\section{Introduction}

\begin{table*}
  \begin{center}
   \caption{Log of Suzaku and XMM-Newton observations.}\label{tab:obs}
    \begin{tabular}{ccccccc}
      \hline
      Telescope/ & Sequence & Field & \multicolumn{2}{c}{Aim point} & Observation & Exposure \\
      Instrument & number & name & $\alpha$ (J2000.0) & $\delta$ (J2000.0) & start date & time (ks)\footnotemark[$*$] \\ \hline
      Suzaku/XIS & 501059010 & north & \timeform{17h48m22s} & \timeform{-27D56'08''} & 2007/03/15 & 62.2/62.2 \\
      Suzaku/XIS & 502020010 & south & \timeform{17h48m46s} & \timeform{-28D08'00''} & 2007/09/06 & 139/139 \\
      XMM-Newton/EPIC & 0112970101 & obs00 & \timeform{17h48m44s} & \timeform{-28D05'06''} & 2000/09/23 & 14.5/12.9 \\
      XMM-Newton/EPIC & 0205240101 & obs05 & \timeform{17h48m17s} & \timeform{-28D07'50''} & 2005/02/26 & 28.7/16.6 \\
      \hline
     \multicolumn{7}{@{}l@{}}{\hbox to 0pt{\parbox{160mm}{\footnotesize
  \footnotemark[$*$] The exposure time after screening of XIS FI/BI for the Suzaku and EPIC MOS/pn for the XMM-Newton observations.
     }\hss}}
    \end{tabular}
  \end{center}
\end{table*}

\begin{figure*}
  \begin{center}
    \FigureFile(180mm,140mm){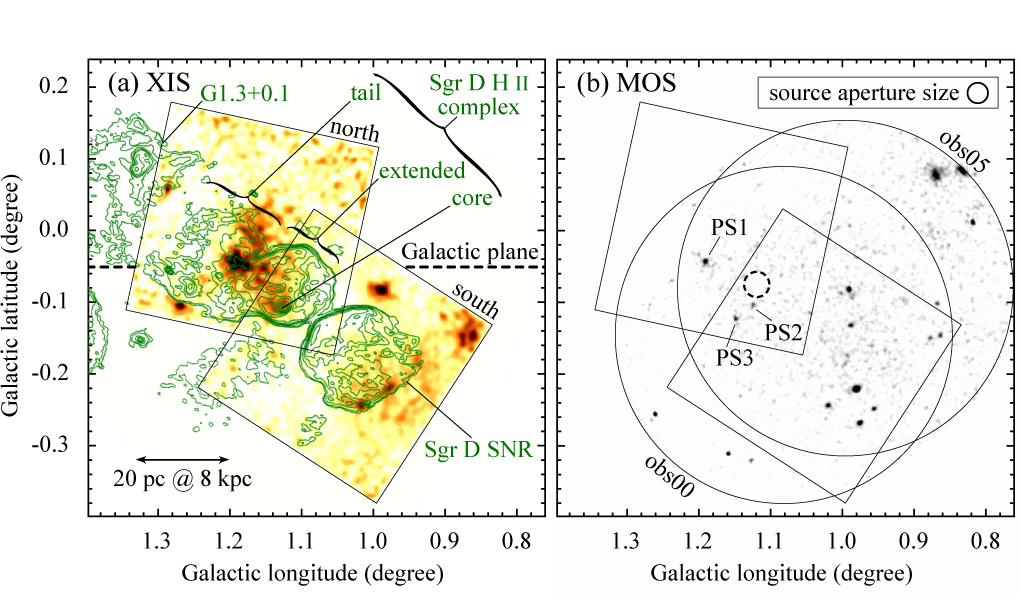}
    %%% \FigureFile(width,height){filename}
  \end{center}
  \caption{Wide-band (0.7--5.5~keV) smoothed images by the (a) XIS and (b) MOS 
shown with the logarithmic intensity scale. Overlaid green contours in 
(a) are an 18~cm radio continuum map from the Very Large Array 
(VLA; \cite{1998ApJ...493..274M}). The point source extraction 
aperture and background accumulation region are shown in (b). 
See \S~\ref{sec:psspec} for details. 
For the XIS image, we merged events with the three CCDs, 
subtracted the non--X-ray background (\cite{2008PASJ...60S..11T}), 
corrected for the vignetting, and mosaicked the two fields 
with different exposure times normalized. 
The fields of view are shown with two solid squares 
with \timeform{18'} in length. For the MOS image, 
we merged two CCDs and mosaicked the two fields with normalized exposures. 
The fields are shown with two circles with \timeform{30'} in diameter.\label{fig:fullimg}}
\end{figure*}

%%%intro. of Sgr D%%%
Sagittarius D (Sgr\,D) is one of the most intense radio sources in the 
Galactic center (GC) region known from the earliest days of 
GC radio observations (\cite{1966ApJ...146..653D}). It is 
composed of an 
H\emissiontype{II} region (Sgr\,D H\emissiontype{II} region or G1.13--0.10) 
and a supernova remnant (SNR; Sgr\,D SNR or G1.05--0.15) adjoining 
each other (figure~\ref{fig:fullimg}a). 
Low-resolution radio continuum observations 
(\cite{Lit1974}; \cite{1974IAUS...60..499S}; \cite{1979A&AS...35....1D}) 
identified thermal emission 
from the H\emissiontype{II} region 
and nonthermal emission from the SNR. 

%%%old obs.%%%
The Sgr\,D H\emissiontype{II} region is associated with its natal 
giant molecular cloud (GMC), 
which has a velocity different from those typical of GMCs 
in the Galactic nuclear disk such as Sgr\,B2 (\cite{1991ApJ...379L..53L}; \cite{1992ApJS...82..495L}). 
In higher resolution radio continuum images (\cite{1998ApJ...493..274M}; \cite{2000AJ....119..207L}; 
\cite{2004ApJS..155..421Y}), 
the Sgr\,D H\emissiontype{II} region is comprised of several components of 
different spatial scales. 
Throughout this paper, we call the entire region ``the Sgr\,D H\emissiontype{II} complex'' 
and treat three components separately (figure~\ref{fig:fullimg}a): 
(1) ``core'', the brightest knot, which is referred to as ``G1.12--0.10'' 
(\cite{1992ApJS...82..495L}; \cite{1998ApJ...493..274M}) or 
the ``compact H\emissiontype{II} region'' (\cite{1999ApJ...512..237B}), 
(2) ``extended'', a $\sim$\timeform{7'} shell surrounding the ``core'' (\cite{1999ApJ...512..237B}), 
and (3) ``tail'', the structure extending eastward from the ``extended'' component, 
which is labeled ``G1.2+0.0'' (\cite{2008ApJS...01..001L}). 
The ``core'' is an H\emissiontype{II} region with a Lyman continuum photon intensity 
equivalent of an O5.5 star based on its infrared properties 
(\cite{1984ApJ...283..601O}; \cite{1999ApJ...512..237B}). 
The latter two structures may be at the edge of 
the H\emissiontype{II} region; 
the ``extended'' component is created by ionizing photons from the ``core'' and 
the ``tail'' by gas escaping away from the ``extended'' component (\cite{1998ApJ...493..274M}). 

%%%recent obs.%%%
Recent observational results, however, 
are diverging away from this view of the 
Sgr\,D H\emissiontype{II} complex. There are two contradictory 
estimates of its distance. 
Narrow molecular line emission and broad H$_{2}$CO absorption features 
indicate that the complex is on the far side of the GC region 
(\cite{1998ApJ...493..274M}), whereas the near-infrared star counts 
around the ``core'' suggest the system 
lies on the near side of the GC (\cite{1999ApJ...512..237B}), possibly in the 
$\sim$160~pc expanding molecular ring (\cite{1991ApJ...379L..53L}).  
The latest radio continuum study 
on arcminute spatial scales indicates a mixture of thermal 
and nonthermal emission in this complex 
\citep{2008ApJS...01..001L}, which historically had been considered solely 
a thermal radio source.

%%%new view of Sgr D HII%%%
These new lines of evidence do not support the long-standing 
interpretation that the Sgr\,D H\emissiontype{II} complex is 
a single H\emissiontype{II} region and its environment. Rather, it is more 
likely that the system is made up of multiple components at various 
distances projected along the same line of sight. Previous results should be 
carefully reviewed and new observations should be conducted to better understand 
the system. 

%%%X-ray diffuse%%%
X-ray observations are indispensable in this pursuit. Both SNRs and
GMCs can be identified through X-ray imaging as emission extended by
$\sim$10~pc. The two types of emission can be distinguished by
their X-ray spectra; SNRs show
thermal plasma with emission lines from highly ionized atoms
(e.g., \cite{1979ApJ...234L..73B}), while GMCs show emission produced 
by external X-ray irradiation (\cite{2000ApJ...534..283M}). 
H\emissiontype{II} regions are also 
an emerging class of diffuse X-ray sources with a similar spatial scale. 
They show a variety of spectral shapes, including soft thermal 
(\cite{2003ApJ...593..874T}; \cite{2008PASJ...60S..85H}), 
hard thermal (\cite{2002ApJ...573..191M}; \cite{2006ApJ...638..860E}), 
and nonthermal (\cite{2002ApJ...580L.161W}; \cite{2004ApJ...611..858L}; 
\cite{2006MNRAS.371...38W}; \cite{2007PASJ...59S.229T}; \cite{2006ApJ...638..860E}) 
emission. In addition to the X-ray emission, the measurement 
of the X-ray absorption gives a constraint on the distance 
and hence the physical scale of extended objects. 

%%%X-ray obs. of Sgr D%%%
Several X-ray observations were reported in the Sgr\,D region. 
In a BeppoSAX study (\cite{2001A&A...372..651S}), diffuse X-ray emission was 
significantly detected from the Sgr\,D SNR and marginally detected from 
the Sgr\,D H\emissiontype{II} complex. 
In an ASCA study (\cite{2002ApJS..138...19S}), the image was plagued
by stray lights from a nearby bright source and was unsuitable to search 
for diffuse X-ray sources. 
In an XMM-Newton study (\cite{2006A&A...456..287S}), 
dozens of point sources were identified, but no diffuse emission was detected 
presumably due to high background. The possible diffuse X-ray
detection by BeppoSAX in the Sgr\,D H\emissiontype{II} complex has not been confirmed 
and no spectral information exists for this source.

%%%Our obs.%%%
We conducted X-ray observations of the Sgr\,D H\emissiontype{II} complex 
using the X-ray Imaging Spectrometer (XIS: \cite{2007PASJ...59S..23K}) 
onboard Suzaku (\cite{2007PASJ...59S...1M}). 
The low background of XIS makes it 
particularly well-suited for finding diffuse sources of low surface brightness and yielding 
their high signal-to-noise ratio spectra. Indeed, a series of XIS studies in the GC region identified 
several new SNRs and irradiated GMCs (\cite{2007PASJ...59S.221K}; 
\cite{2008PASJ...60S.183M}; \cite{2008PASJ...60S.191N})
 and reported detailed spectroscopy of an H\emissiontype{II} region 
(\cite{2007PASJ...59S.229T}). Upon the confirmation of
the previously claimed marginal diffuse detection in the Sgr\,D H\emissiontype{II} complex, 
we further aim to construct the X-ray spectrum 
which gives important insights into the origin of 
the emission and the entire complex.

\medskip

%%%Our work%%%
Here, we present a significant detection of diffuse X-ray emission from
the Sgr\,D H\emissiontype{II} complex with the Suzaku XIS. 
High signal-to-noise ratio spectra were obtained from two different 
diffuse sources. 
We discuss their X-ray characteristics and their association with 
sources observed in our study using the 100~m Green Bank Telescope (GBT) 
and in other archived multiwavelength data sets. 
Based on these data, we propose a new view of the Sgr\,D H\emissiontype{II} 
complex. 
In this paper, we supplement the Suzaku data 
with those taken by XMM-Newton in order to evaluate 
the contribution of point sources to the diffuse emission. 
Throughout this paper, we use east and west in the Galactic longitude direction, 
and north and south in the Galactic latitude direction for simplicity.

\section{Observations}

\subsection{Suzaku}
%%%Suzaku observation of Sgr D%%%
We used two XIS fields covering the Sgr\,D H \emissiontype{II} complex in the 
Suzaku GC mapping project (table~\ref{tab:obs}). 
We hereafter call the two fields as north and south 
field (figure~\ref{fig:fullimg}a). 
The north field covers the entire complex. 
The overlapping area of the north and south fields also cover a part of 
the complex, for which we additionally use the south field data. 
Results in the remaining part of the south field will be reported in a separate paper.

%%%introduction of the XIS%%%
The XIS consists of four CCD cameras (XIS 0, 1, 2, and 3) 
each placed at the focal planes of the four X-ray Telescopes 
(XRT: Serlemitsos et al. 2007).
One of the cameras (XIS~1) uses a back-illuminated (BI) CCD 
while the others (XIS~0, 2, and 3) use front-illuminated (FI) CCDs. 
One of the FIs (XIS~2) has not been functional since an unexpected 
anomaly in 2006 November. 

%%%performance of the XIS and the XRT%%%
%%spectroscopy properties%%
The XIS is sensitive in an energy range of 0.2--12~keV with 
the FI CCDs superior in the hard band and the BI CCD in the soft band.  
For the purpose of mitigating the degradation of the energy resolution in the orbit, 
the XIS employs the spaced-row charge injection (SCI) technique 
(\cite{2004SPIE.5501..111B}; \cite{2007SPIE.6686...U}). 
Artificial charges are periodically injected to fill 
charge traps caused by radiation damage in order to 
reduce the charge transfer inefficiency (CTI). 
In the present data, the resolution at 5.9~keV is 
measured to be $\sim$150~eV (FI) and $\sim$170~eV (BI) in the 
full width at half maximum (FWHM) using the $^{55}$Fe calibration sources 
installed in the two corners of each CCD. 

%%imaging properties%%
%size of FOV etc.%
Combined with the XRT, an \timeform{18'}$\times$\timeform{18'} 
region is covered with a pixel scale of \timeform{1''}~pixel$^{-1}$. 
An angular resolution of \timeform{1.9'}--\timeform{2.3'} 
in the half power diameter (HPD) is almost independent of 
the off-axis angles within $\sim$10$\%$. 
%effective area and background environment%
The total effective area excluding the inoperational XIS~2 is 
$\sim$430~cm$^2$ at 8~keV. 
Due to the Suzaku's low orbital altitude at $\sim$550~km, 
the XIS achieves a low and stable background environment 
suited for studying faint diffuse sources.

%%%screening%%%
In the two observations presented here, the XIS was operated 
in the normal clocking mode with the SCI technique. 
Starting from the pipeline products in the processing version 2.0, 
we recalculated the energy scales by taking into account of 
the time dependence of the CTI and the positional dependence 
of the SCI efficiency.\footnote{See 
(http://heasarc.gsfc.nasa.gov/docs/suzaku/analysis/sci\_gain\\\_update.html) 
for details.} The resultant systematic uncertainty in the energy gain 
is $\lesssim$10~eV at 5.9~keV.
We removed events during the South Atlantic Anomaly passages, 
Earth elevation angles below \timeform{5D}, 
and Earth day-time elevation angles below \timeform{20D}. 
We also removed hot and flickering pixels. 
The net exposures are 62.2 and 139~ks for the north and 
south fields, respectively. 
We reduced the data using the software packages 
HEADAS\footnote{See (http://heasarc.nasa.gov/lheasoft/).} 
version 6.4 and XSpec \citep{1996ASPC..101...17A} version 11.3.2. 
In the spectral analysis, we used the redistribution matrix function 
released on 2006-02-13 
%that best reproduces the observed calibration source spectra 
and the auxiliary response function generated 
by a ray-tracing simulator (\texttt{xissimarfgen}: \cite{2007PASJ...59S.113I}).

\subsection{XMM-Newton}
%%%XMM-Newton observation of Sgr D%%%
We retrieved the archived XMM-Newton \citep{2001A&A...365L...1J} data 
to supplement the Suzaku data, and found that two observations taken 
in 2000 and 2005 cover Sgr\,D. We hereafter refer to them 
as obs00 and obs05, respectively (table~\ref{tab:obs}). 
The obs05 data were presented by \citet{2006A&A...456..287S}, upon 
which we add the obs00 data to make deeper images. 
Both data sets were obtained with the European Photon Imaging Camera (EPIC), 
which is comprised of two MOS CCDs \citep{2001A&A...365L..27T} 
and a pn CCD \citep{2001A&A...365L..18S}. 

%%spectroscopy properties%%
%energy resolution etc.%
The EPIC is sensitive in an energy range of 0.1--10~keV 
with an energy resolution of $\sim$140~eV (MOS) and $\sim$170~eV (pn) in the 
FWHM at 5.9~keV for the retrieved data. 
The MOS and pn cover a $\sim$\timeform{30'} diameter circle with a pixel scale of 
\timeform{1''}~pixel$^{-1}$ and \timeform{4''}~pixel$^{-1}$, respectively. 
The HPD and the total effective area at 8~keV are 
\timeform{4.2''}--\timeform{6.6''} at the optical axis and $\sim$900~cm$^2$, respectively. 
The combination of XMM-Newton's large effective area and imaging capability 
makes EPIC observations highly sensitive to point sources.

%%%property of the obs., screening etc.%%%
In the two observations, both the MOS and the pn were operated 
in the full frame mode with the medium filter. 
We removed events taken during periods with 
background rates higher than $0.5$~counts~s$^{-1}$ (MOS) 
and $0.3$~counts~s$^{-1}$ (pn) 
in the 10--15~keV energy band. 
We also removed hot and flickering pixels. 
The resultant effective exposures are shown in table~\ref{tab:obs}. 
The XMM-Newton Science Analysis System\footnote{See (http://xmm.vilspa.esa.es/sas/).} 
version 7.1.0 was used for reducing the EPIC data.

\begin{figure*}
  \begin{center}
    \FigureFile(180mm,200mm){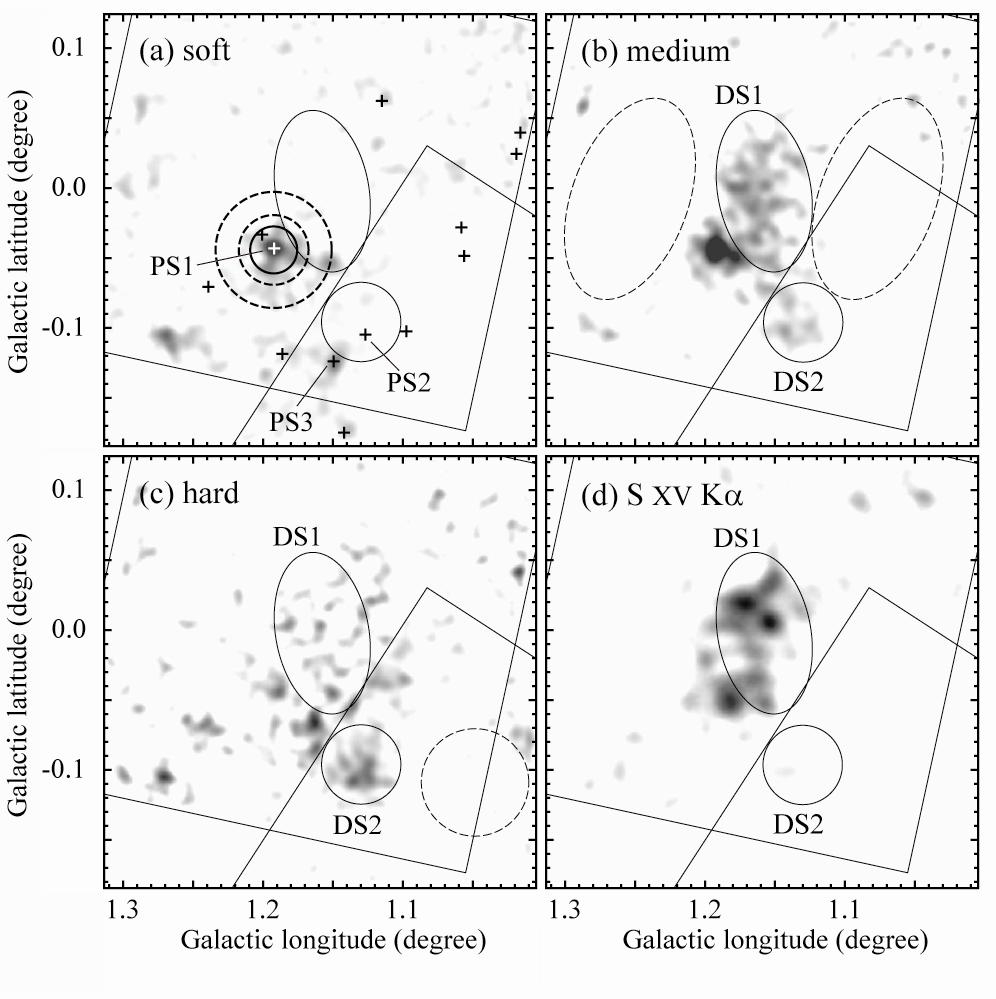}
    %%% \FigureFile(width,height){filename}
  \end{center}
  \caption{Close-up XIS images of the Sgr\,D H \emissiontype{II} complex 
in various energy bands: (a) soft (0.7--2.0~keV), (b) medium (2.0--4.0~keV), 
(c) hard (4.0--5.5~keV), and (d) S \emissiontype{XV} K$\alpha$ (2.4--2.5~keV). 
Crosses in (a) represent positions of point sources by XMM-Newton \citep{2007A&G....48e..30W}. 
The source and background extraction regions are shown 
with solid and broken symbols, respectively; thick circles for PS1 
and thin ellipses for DS1 and DS2. See \S~\ref{sec:psspec} and \S~\ref{sec:dsspec} for details. 
The images were processed in the same way as the wide-band XIS image 
(figure~\ref{fig:fullimg}a). \label{fig:xisimg}}
\end{figure*}

\section{Results}

\subsection{Images}

\subsubsection{Wide-Band Images}
%%%About images and overall features%%%
Figure~\ref{fig:fullimg} shows the wide-band (0.7--5.5~keV) 
X-ray images obtained with the XIS and the EPIC MOS. 
Here, we restricted the energy range to be below 5.5~keV to 
eliminate the signals from the calibration sources in the XIS image. 
We find that the Sgr\,D H \emissiontype{II} complex is accompanied by 
both diffuse and point-like X-ray emission. 
The former is more noticeable in the XIS image while the latter is in 
the MOS image, reflecting the complementary capability of the two imagers.

%%%PS in MOS and Astrometry%%%
In the MOS image (figure~\ref{fig:fullimg}b), we found three bright 
point-like sources in and around the complex listed in the 
Second XMM-Newton Serendipitous Source Catalogue \citep{2007A&G....48e..30W}: 
PS1 (2XMM~J174835.9--275619), 
PS2 (2XMM J174841.1--280136), and PS3 (2XMM J174848.7--280100). 
We matched the astrometric positions between the XIS and the MOS images 
using bright and isolated point-like sources. 
The absolute astrometry of XMM-Newton is accurate to 
$\sim$\timeform{2''}\footnote{See 
(http://xmm.esac.esa.int/docs/documents/CAL-TN-0018.pdf) for details.}, 
while that of Suzaku can be uncertain up to $\sim$\timeform{20''} 
\citep{2008PASJ...60S..35U}. 
The XIS south field was shifted to match the MOS image by 
($\Delta$R.\,A., $\Delta$Dec.) = (\timeform{0''}, \timeform{22''}), 
while the north field was unchanged as it is consistent with the MOS 
astrometric positions. 

\subsubsection{Band-Limited Images}
%%%XIS broad images%%%
XIS images were exposed for much longer times than MOS images 
(table~\ref{tab:obs}). Sufficient photon counts enable us to construct 
both broad-band and narrow-band images with XIS. 
Figure~\ref{fig:xisimg} shows the close-up view of 
the Sgr\,D H \emissiontype{II} complex in four limited energy bands: 
(a) soft (0.7--2.0~keV), (b) medium (2.0--4.0~keV), (c) hard (4.0--5.5~keV), 
and (d) S \emissiontype{XV} K$\alpha$ (2.4--2.5~keV).

%%%broad-band images%%%
In the three broad-band images 
(figure~\ref{fig:xisimg}a--\ref{fig:xisimg}c), 
PS1 appears only in the 
soft and medium bands, indicating that it is soft, 
lightly attenuated, and thus a foreground source. 
The soft band image lacks diffuse features. 
However, in the medium and hard bands, diffuse emission emerges 
with different colors at the north field center and 
at the Sgr\,D H\emissiontype{II} core. 
We consider that the two diffuse sources are different 
and call them DS1 and DS2 
(figure~\ref{fig:xisimg}). 

%%%narrow-band image%%%
In the narrow-band image tracing the S \emissiontype{XV} 
K$\alpha$ line emission (figure~\ref{fig:xisimg}d), 
PS1 and DS1 are conspicuous. 
These sources are likely to be thermal emission of 
a few~keV plasma temperatures. 
DS2, on the contrary, is faint in this band. 
This reinforces the claim that DS1 and DS2 have different origins. 

\subsection{Spectra}

\subsubsection{Point-Like Sources}\label{sec:psspec}
\begin{figure}
  \begin{center}
    \FigureFile(80mm,160mm){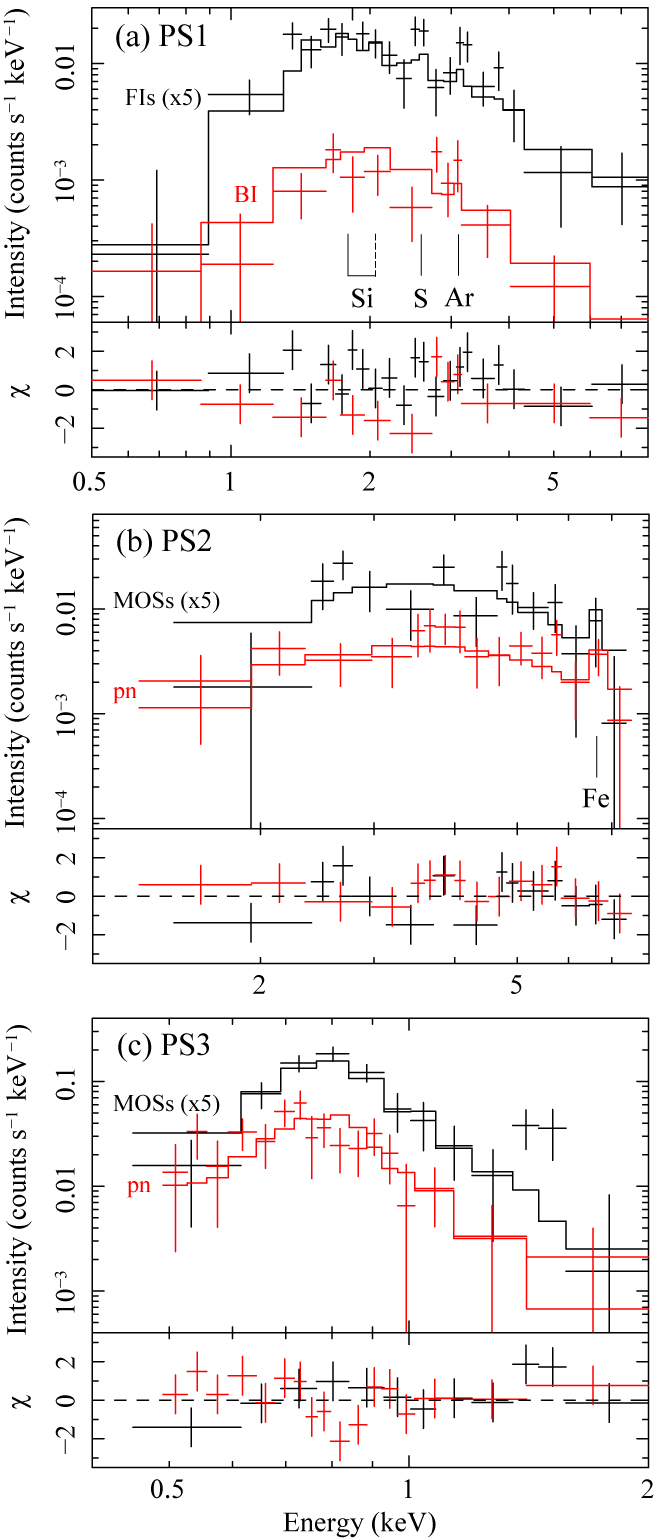}
  \end{center}
  \caption{Background-subtracted spectra (pluses) and the best-fit models 
(histograms) of PS1--3 (a--c) in the top panels and the residuals to the fit 
in the bottom panels. For PS1, XIS merged FI and BI spectra are shown 
in black and red. For PS2 and PS3, EPIC merged MOS and pn spectra are shown 
in black and red. Black spectra are scaled upward by a factor of 5 for clarity.}\label{fig:fitps}
\end{figure}

%%%analyzed data%%%
We first present the spectral fits of the three bright point sources 
that may contaminate the diffuse emission. 
Figure~\ref{fig:fitps} shows their background-subtracted spectra. 
%PS1%
For PS1, we used the XIS data as it is detected and the 
statistics is higher than the EPIC data. 
The source photons were accumulated from a \timeform{1'} radius 
(the thick solid circle in figure~\ref{fig:xisimg}a) to maximize 
the ratio against the background. The background photons were 
from a concentric annulus around the source (the thick dashed annulus) 
to suppress the contamination from DS1. 
In the spectrum (figure~\ref{fig:fitps}a), emission lines are seen 
from several species of highly ionized atoms. We fitted the data with 
a thin-thermal plasma model 
(\texttt{apec} in the XSpec fitting package: \cite{2001ApJ...556L..91S}) 
attenuated by interstellar extinction (\texttt{wabs}: \cite{1983ApJ...270..119M}). 
The chemical abundance is fixed to the solar value \citep{1989GeCoA..53..197A}.
The best-fit parameters are given in table~\ref{tab:pntpar} 
for the amount of extinction ($N_{\rm{H}}$), the plasma temperature ($k_{\rm{B}}T$), 
the X-ray photon flux ($f_{\rm{X}}$), 
and the absorption-corrected X-ray luminosity ($L_{\rm{X}}$). 
We obtained a similar but less constrained result for the 
EPIC spectrum of PS1. 

For PS2 and PS3, we used the EPIC data as they are not detected 
in the XIS image. Source photons were extracted from a 90\% 
encircled energy radius of $\sim$\timeform{50''} for both sources 
(the aperture size is shown at the top right in figure~\ref{fig:fullimg}b). 
The background common to both sources was taken from the region shown 
with the thick dashed circle in figure~\ref{fig:fullimg}b. 
A suggestion of a faint Fe K emission line is found for PS2 in the EPIC spectra, 
but other lines are not seen due to lack of counts 
(figure~\ref{fig:fitps}b and \ref{fig:fitps}c). 
The PS2 and PS3 spectra were fitted using the same model with PS1. 
The best-fit parameters are also given in table~\ref{tab:pntpar}. 

\begin{table}
 \begin{center}
  \caption{Best-fit parameters of the point-like sources.}\label{tab:pntpar}
%   \begin{minipage}{80mm}
   \begin{tabular}{lccc}
      \hline
      Parameter & PS1 & PS2 & PS3 \\ \hline
      Data & XIS & EPIC & EPIC \\ \hline
      $N_{\rm{H}}$ (10$^{22}$~cm$^{-2}$)\footnotemark[$*$] & 2.1$^{+0.6}_{-0.5}$ & 5.7$^{+2.0}_{-1.6}$ & $<$0.30 \\
      $k_{\rm{B}}T$ (keV)\footnotemark[$*$] & 1.9$^{+0.6}_{-0.4}$ & 7.6$^{+\infty}_{-7.6}$ & 0.48$^{+0.08}_{-0.12}$ \\
      $f_{\rm{X}}$ (s$^{-1}$~cm$^{-2}$)\footnotemark[$\dagger$] & 2.9$\times 10^{-5}$ & 1.7$\times 10^{-5}$ & 1.3$\times 10^{-5}$ \\ 
      $L_{\rm{X}}$ (erg~s$^{-1}$)\footnotemark[$\dagger$]\footnotemark[$\ddagger$] & 7.4$\times 10^{32}$ & 2.1$\times 10^{33}$ & $<$2.2$\times 10^{30}$ \\
      $\chi ^2$/d.o.f. & 46/32 & 24/27 & 26/26 \\ \hline
       \multicolumn{4}{@{}l@{}}{\hbox to 0pt{\parbox{85mm}{\footnotesize
       \par\noindent
       \footnotemark[$*$] Statistical uncertainty is represented by the 90\% confidence intervals. 
       \par\noindent
       \footnotemark[$\dagger$] Values in the 0.7--8.0~keV band. 
       \par\noindent
       \footnotemark[$\ddagger$] The absorption is corrected and the distances are assumed to be 6.8~kpc (PS1), 
                                8.0~kpc (PS2), and $<$0.97~kpc (PS3). See \S\ref{sec:disothers} for details.

     }\hss}}
    \end{tabular}
%   \end{minipage}
  \end{center}
\end{table}

\subsubsection{Diffuse Sources}\label{sec:dsspec}
\begin{figure}
  \begin{center}
    \FigureFile(80mm,100mm){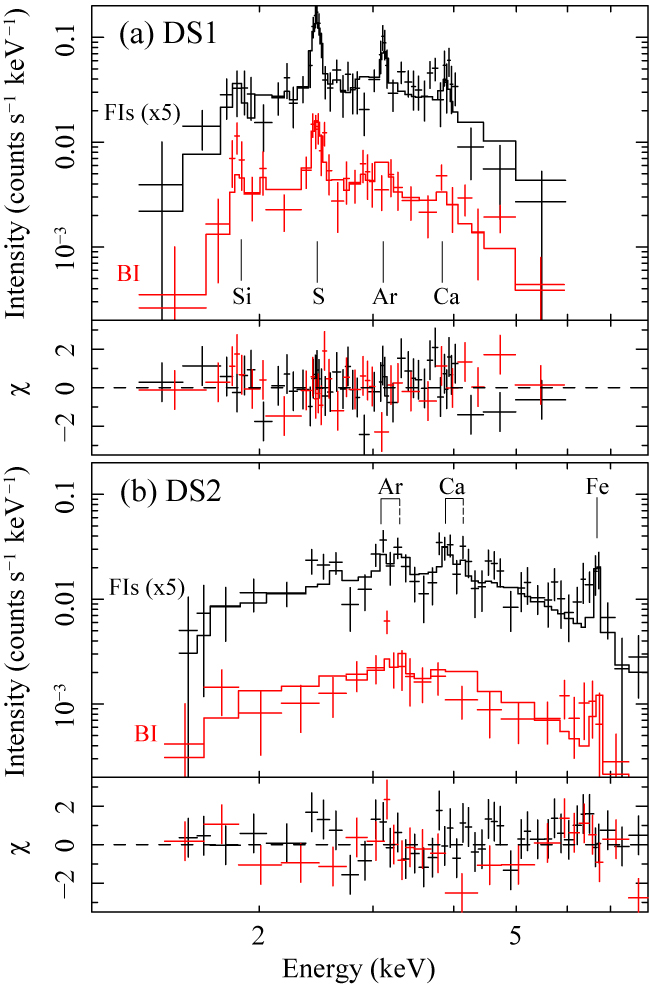}
  \end{center}
  \caption{Background-subtracted XIS spectra and the best-fit models of 
(a) DS1 and (b) DS2. Symbols follow figure~\ref{fig:fitps}.}\label{fig:fitds}
\end{figure}

%%%DS1%%%
We next present the spectra of diffuse sources using XIS. 
Figure~\ref{fig:fitds} shows their background-subtracted spectra. 
For DS1, source photons were extracted from a region defined by the emission 
morphology in the S \emissiontype{XV} narrow-band image 
(the thin solid ellipse in figure~\ref{fig:xisimg}). 
Spectra of faint diffuse sources are prone to uncertainties 
stemming from the background subtraction. 
We considered the following to minimize possible systematic effects. 
First, the data in the same observation (the north field) were used both for 
the source and background signals. Second, the background spectrum was constructed 
from two regions at the same Galactic latitude 
(the thin dashed ellipses in figure~\ref{fig:xisimg}b) as the background 
is dominated by the Galactic center diffuse X-ray emission distributed symmetrically 
with respect to the Galactic plane (\cite{1989Natur.339..603K}; \cite{1990ApJ...365..532Y}). 
Third, the difference in the effective area due to 
the different off-axis angles between the source and background regions 
was taken into account (\cite{2008PASJ...60S..85H}).

%%%DS2%%%
For DS2, we used both the north and south field data. 
The source spectrum was integrated from a $\sim$\timeform{1.7'} radius 
to encompass the Sgr\,D H\emissiontype{II} core (the thin solid circle in figure 
\ref{fig:xisimg}). 
The background  subtraction was performed separately in each field and 
then merged. 
In the north field, we utilized the background region 
for DS1, which happens to be at the symmetric position to the 
Galactic plane with DS2. In the south field, the background was 
constructed from the thin dashed circle in figure~\ref{fig:xisimg}c. 
Other procedures follow DS1. 

%%%fits%%%
The spectra of DS1 and DS2 are both characterized by emission lines 
from highly ionized atoms (figure \ref{fig:fitds}). 
Thermal fits were tried using the same model 
with the point sources. Because of stronger signal, 
we are able to derive the chemical abundances relative to solar for elements 
with conspicuous emission lines ($Z_{\rm{S}}$, $Z_{\rm{Ar}}$, 
$Z_{\rm{Ca}}$, and $Z_{\rm{Fe}}$). 
The best-fit values are summarized in table~\ref{tab:specpar}. 
For DS1, we also tried a non-equilibrium ionization model (\texttt{nei}; 
\cite{2001ApJ...548..820B}). This examines how much a plasma 
has relaxed toward a collisional equilibrium among ions and electrons. 
The degree of the relaxation is parameterized by $n_{\rm{e}}t$, 
where $n_{\rm{e}}$ is the electron density and $t$ is the time from 
the plasma creation. A lower limit of $n_{\rm{e}}t$ was 
derived as $\sim$10$^{11.2}$~s~cm$^{-3}$, which is consistent with 
DS1 being in collisional equilibrium 
($n_{\rm{e}}t>$10$^{12}$~s~cm$^{-3}$; \cite{1984Ap&SS..98..367M}). 
For DS2, we found that an additional 
Gaussian model at $\sim$6.4~keV improves the fit 
(F-test significance of $\sim$98\%) with an equivalent width of $\sim$270~eV. 
More counts are necessary to confirm this. 

%%%time variability%%%
Because the DS2 emission may largely originate from unresolved point sources, we 
examined time variation. Light curves in several energy bands 
are tested against the null hypothesis that the flux is constant using the 
$\chi^{2}$ statistics. We found a marginal variability 
(significant at a $\sim$97\% level) in DS2 
in the north field data. For comparison, we conducted the same 
procedure for DS1 and the background regions. Their signals were found to be
constant at high confidence levels.  We thus believe that the procedure 
to test the time variability is reliable, and that DS2 varies. 

\begin{table}
  \begin{center}
   \caption{Best-fit parameters of the diffuse sources by XIS.}\label{tab:specpar}
   \begin{tabular}{lcc}
      \hline
      Parameter & DS1 & DS2 \\ \hline
      $N_{\rm{H}}$ (10$^{22}$ cm$^{-2}$)\footnotemark[$*$] & 8.5$^{+1.4}_{-1.2}$ & 5.6$^{+2.3}_{-1.0}$ \\
      $k_{\rm{B}}T$ (keV)\footnotemark[$*$] & 0.91$^{+0.17}_{-0.17}$ & 3.4$^{+1.1}_{-1.2}$ \\
      $Z_{\rm{S}}$ ($Z_{\odot}$)\footnotemark[$*$] & 1.6$^{+0.5}_{-0.4}$ & 1 (fixed) \\
      $Z_{\rm{Ar}}$ ($Z_{\odot}$)\footnotemark[$*$]\footnotemark[$\dagger$] & 1.8$^{+1.1}_{-0.9}$ & 4.8$^{+4.0}_{-3.0}$ \\
      $Z_{\rm{Ca}}$ ($Z_{\odot}$)\footnotemark[$*$]\footnotemark[$\dagger$] & 1.8$^{+1.1}_{-0.9}$ & 4.8$^{+4.0}_{-3.0}$ \\
      $Z_{\rm{Fe}}$ ($Z_{\odot}$)\footnotemark[$*$] & 1 (fixed) & 0.52$^{+0.41}_{-0.28}$ \\
      $f_{\rm{X}}$\footnotemark[$\ddagger$] (s$^{-1}$~cm$^{-2}$) & 6.9$\times 10^{-5}$ & 4.1$\times 10^{-5}$ \\
      $L_{\rm{X}}$\footnotemark[$\ddagger$]\footnotemark[$\S$] (erg~s$^{-1}$) & 1.4$\times 10^{35}$ & 6.0$\times 10^{33}$ \\
      $\chi^2$/d.o.f. & 66/75 & 70/65 \\ \hline
       \multicolumn{3}{@{}l@{}}{\hbox to 0pt{\parbox{70mm}{\footnotesize
       \par\noindent
       \footnotemark[$*$] Statistical uncertainty is represented by 
       the 90\% confidence intervals. 
       \par\noindent
       \footnotemark[$\dagger$] The values of $Z_{\mathrm{Ar}}$ and $Z_{\mathrm{Ca}}$ are fixed to one another. 
       \par\noindent
       \footnotemark[$\ddagger$] Values in the 0.7--8.0~keV band. 
       \par\noindent
       \footnotemark[$\S$] The absorption is corrected and the distances are assumed to be 8.0~kpc. 
           See \S~\ref{sec:disds1} and \S~\ref{sec:disothers} for details.
     }\hss}}
   \end{tabular}
  \end{center}
\end{table}

\section{Discussion}

\subsection{Diffuse Source DS1}\label{sec:disds1}

\subsubsection{Nature of X-ray Emission}
We first argue that the level of contamination to DS1 by point sources is
negligible. The largest contaminant is PS1 among all point sources listed in the XMM-Newton 
catalog (pluses in figure 2a), but its contribution accounts only for $\sim$3\% of the
total XIS counts of DS1. Other point sources make even smaller contributions.

We next argue that DS1 is an extended source. 
We retrieved optical images using the Digitized Sky Survey from the Space Telescope Science 
Institute\footnote{See (http://archive.stsci.edu/cgi-bin/dss\_form) for details.} and 
near-infrared (NIR) and mid-infrared (MIR) images respectively using the Two Micron All Sky
Survey (\cite{1997ASSL..210...25S}) and the Spitzer Space Telescope 
(\cite{2004ApJS..154....1W}) from the Infrared Science 
Archive\footnote{See (http://irsa.ipac.caltech.edu/) for details.}. 
We compared these images with our X-ray image and found no stellar distribution 
with a morphology similar to DS1. 
The point source distribution in the EPIC image (figure 1b) 
differs significantly from the surface brightness distribution 
of DS1 in the XIS image (figure~\ref{fig:fullimg}a). 
We thus conclude that DS1 is extended in nature, and not an ensemble of unresolved point sources.

The extinction ($\sim$8.5$\times$10$^{22}$~cm$^{-2}$) derived from the X-ray spectral fit 
(table~\ref{tab:specpar}) indicates that DS1 is in the GC region 
(\cite{2003ApJ...591..891B}). 
The apparent extent converts to $\sim$9$\times$16~pc at an 8~kpc distance \citep{1993ARA&A..31..345R}. 
A source like DS1 has not been previously reported in the literature. 
It is therefore a new celestial object.

\subsubsection{Origin of X-ray Emission}
Classes of diffuse X-ray emission with an extent of $\sim$10~pc include 
SNRs, irradiated GMCs, pulsar wind nebulae (PWN; \cite{2008ApJ...673..251M}), and 
H\emissiontype{II} regions (\cite{2002ApJ...570..665Y}; \cite{2004ApJ...611..858L}; 
\cite{2006MNRAS.371...38W}; \cite{2007PASJ...59S.229T}) in the GC region. 
Since DS1 shows unambiguous thermal features of a $\sim$0.9~keV temperature with 
enhanced abundances of metals, the GMC and PWN origins are unlikely.
At the position of DS1, H\emissiontype{II} regions are not known 
nor found in our inspection of Spitzer MIR images, which makes                         
an H\emissiontype{II} region origin also unlikely. 
The XIS spectrum is akin to those
found in SNRs in the GC region (\cite{2007PASJ...59S.221K}; 
\cite{2008PASJ...60S.183M}; \cite{2008PASJ...60S.191N}). 
Thus, we propose that DS1 is X-ray emission from a new SNR. 
The lack of X-ray variation is consistent with the SNR interpretation. 

SNRs are expected to show nonthermal synchrotron emission
in centimeter bands. The Sgr\,D H\emissiontype{II} complex has long 
been considered to have
thermal emission, not nonthermal emission, from the flat radio spectral indices 
(\cite{1966ApJ...146..653D}; \cite{1974IAUS...60..499S}; 
\cite{Lit1974}; \cite{1979A&AS...35....1D}). 
However, all these results were based on low resolution data, and the spectral indices
were dominated by two H\emissiontype{II} regions with strong thermal emission in 
and around this complex; 
the Sgr\,D H\emissiontype{II} core and the other H\emissiontype{II} region 
to the east (G1.3+0.1 in figure~\ref{fig:fullimg}a; \cite{1998ApJ...493..274M}). 

The latest work by \citet{2008ApJS...01..001L} presented radio continuum maps 
on spatial scales comparable to the X-ray images. 
The spectral indices were derived from the 3.5~cm and 6.0~cm continuum maps 
obtained with GBT along several slices across the Sgr\,D H\emissiontype{II} 
complex. Some indices indicate that there is a mixture of 
thermal and nonthermal emission. 

\begin{figure}
  \begin{center}
    \FigureFile(85mm,200mm){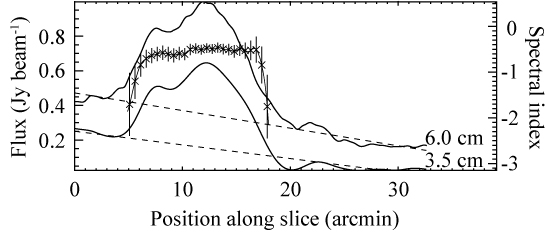}
    %%% \FigureFile(width,height){filename}
  \end{center}
  \caption{Profiles of 3.5~cm and 6.0~cm intensity (solid curves) and the spectral index (crosses) between the two bands along the slice shown in figure~\ref{fig:gbt}a. Background (dashed lines) is derived by linearly fitting the slope at the two sides of the source region, where the indices are computed. The error bars indicate 1$\sigma$ uncertainty.} \label{fig:slice}
\end{figure}

Using this GBT data set, we constructed the intensity and the spectral index 
profiles along a slice including DS1 but not the Sgr\,D H\emissiontype{II} core 
(figure~\ref{fig:slice}). The slice was taken from two images (3.5~cm and 6.0~cm) 
convolved to the same beam size of $\sim$\timeform{2.5'}. 
Details of the data and other reduction procedures are described in \citet{2008ApJS...01..001L}. 
The background-subtracted index is consistently about --0.5, 
indicating that nonthermal synchrotron emission dominates the 
spectrum (\cite{2004BASI...32..335G}). The radio continuum peak corresponds to 
the position of DS1. This result constitutes independent evidence for classifying DS1 
as an SNR.

\begin{figure*}
  \begin{center}
    \FigureFile(180mm,200mm){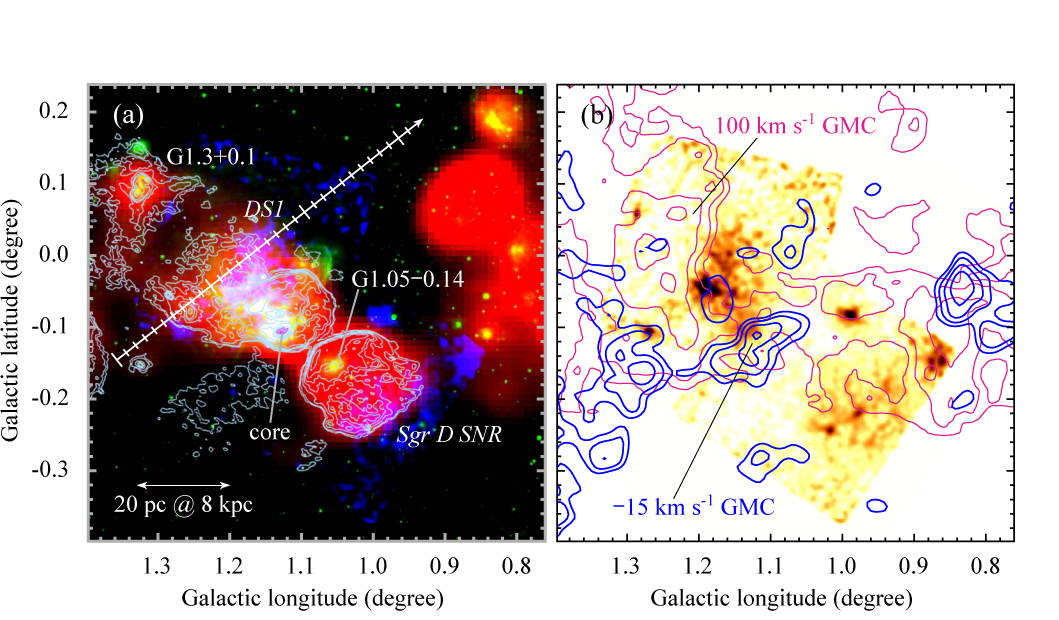}
    %%% \FigureFile(width,height){filename}
  \end{center}
  \caption{(a) Composite color image with Suzaku X-ray (0.7--5.5~keV) in blue, Spitzer MIR (24~$\mu$m) in green, and GBT radio (6.0~cm) in red. The slice for the radio continuum index (figure~\ref{fig:slice}) is shown with a ticked vector. Objects are labeled in Italic for SNRs and in Roman for H\emissiontype{II} regions. (b) X-ray image with CO ($J=$~3--2) emission at 100$\pm$5~km~s$^{-1}$ (\cite{2007PASJ...59...15O}) in red contours and CS ($J=$1--0) emission at $-$15$\pm$5~km~s$^{-1}$ (\cite{1999ApJS..120....1T}) in blue contours.} \label{fig:gbt}
\end{figure*}

\subsubsection{Structure of the Sgr\,D~H\emissiontype{II} Complex}
We now have sufficient information to decompose and deproject the Sgr\,D H\emissiontype{II} 
complex. The data are displayed in figure~\ref{fig:gbt} with the same scale as 
figure~\ref{fig:fullimg}. Figure~\ref{fig:gbt} (a) is composed of the Suzaku X-ray (0.7--5.5~keV; blue), 
Spitzer MIR (24~$\mu$m; green), and GBT radio continuum (6.0~cm; red) images. In this color 
code, SNRs appear in magenta with X-ray and nonthermal radio emission such as DS1 and the southwestern part 
of the Sgr\,D SNR. H\emissiontype{II} regions appear in yellow with MIR and thermal radio emission, which 
include G1.3+0.1 and another H\emissiontype{II} region inferred by \citet{1979A&AS...35....1D} 
in the northeast of the Sgr\,D SNR (we name this component G1.05--0.14). The Sgr\,D H\emissiontype{II} core, 
which has emission in all three bands, does not 
appear white because the Spitzer image is saturated. 
Figure~\ref{fig:gbt} (b) is a superposition of the X-ray image with millimeter emission line maps in two different 
velocities at the local standard of rest; 100~km~s$^{-1}$ by CO ($J=$~3--2) in red contours 
(\cite{2007PASJ...59...15O}) taken by the Atacama Submillimeter Telescope Experiment and $-$15~km~s$^{-1}$ by CS ($J=$~1--0) in blue contours 
(\cite{1999ApJS..120....1T}) taken by the Nobeyama 45~m telescope. 

We argue that ``tail'' is the new SNR (we name G1.2--0.0) and the northwestern part of it is bright in X-ray as 
DS1. The X-ray morphology of DS1 is not well-correlated with the well-defined shell labeled 
``extended'' in figure~\ref{fig:fullimg}a. It is therefore unlikely that the two objects 
are physically related. It is more reasonable to consider that DS1 is associated with ``tail''. 
The index across ``tail'' is almost constant (figure~\ref{fig:slice}), 
suggesting that this is a uniform structure. Both 
X-ray and radio properties indicate that this is an SNR. The X-ray is bright only 
in the western part of the imperfect shell-like morphology of the ``tail''. 
The lack of X-ray emission in other parts is attributable to an intervening 
GMC seen in figure~\ref{fig:gbt} (b). The CO emission with a 100~km~s$^{-1}$ 
velocity (red contours) is anti-correlated with the X-ray emission; DS1 is just 
outside of the sharp boundary of the CO cloud. The extinction along this GMC 
is $N_{\rm{H}}\sim$10$^{24}$~cm$^{-2}$ from the CO data (private communication with T. Oka), 
which is sufficient to block the $k_{\rm{B}}T\sim$0.9~keV X-ray emission behind it.

The Sgr\,D H\emissiontype{II} core is associated with a GMC in 
a different velocity (blue contours), which was suggested by \citet{1991ApJ...379L..53L} 
and \citet{1992ApJS...82..495L}. \citet{1999ApJ...512..237B} proposed that ``core'' is a 
blister H\emissiontype{II} region on the GMC. We speculate that the H\emissiontype{II} 
region G1.05--0.14 is another blister on the other side of the GMC. 
H~109$\alpha$ emission was detected at this position and its velocity is 
the same with the GMC (\cite{1975A&A....44..259P}).

It is now clear that Sgr\,D H\emissiontype{II} complex is not a single H\emissiontype{II} 
region and its environment as was considered in the past. 
We propose that an SNR and two H\emissiontype{II} regions on both sides of a GMC are 
projected along the same line of sight. The SNR lies behind the H\emissiontype{II} regions, which is 
justified by a larger $N_{\rm{H}}$ value in DS1 (a part of the SNR) than in DS2 (``core''). 
The SNR position behind the 100~km~s$^{-1}$ cloud puts this object in or behind the GC region, 
whereas the H\emissiontype{II} regions associated with a GMC at $-$15~km~s$^{-1}$ is 
on the near side of the GC in agreement with the NIR star count study (\cite{1999ApJ...512..237B}).

\subsection{Other X-ray Sources}\label{sec:disothers}
\subsubsection{Point Sources PS1--PS3}
PS1 and PS3 are sources 7 and 11 respectively in the XMM-Newton study by \citet{2006A&A...456..287S}. 
PS1 has several
anonymous NIR sources within the positional uncertainty, and its counterpart 
cannot be identified. 
PS3 is identified as HD\,316290 in \citet{2006A&A...456..287S}, which is an F8 
star with the \textit{B}-band and \textit{V}-band magnitudes 
of 10.3~mag and 9.8~mag, respectively (\cite{1995A&AS..110..367N}).

We consider that both sources are foreground objects from their values of
$N_{\rm{H}}$. Assuming an interstellar hydrogen density of $n_{\rm{H}}=$~1~cm$^{-3}$, we
derived the distance to PS1 as $\sim$6.8 kpc and constrained that to PS3 
as $\lesssim$0.97 kpc, respectively. 
The X-ray spectrum and luminosity suggest that both
sources are individual stars, but their nature is not constrained any further.

PS2, on the other hand, shows an $N_{\rm{H}}$ value consistent with being in the
GC region. We identify a bright NIR source (star A in \cite{1999ApJ...512..237B}) within 
the position uncertainty of PS2, which is located $\sim$8\arcsec\ north to the center of 
the Sgr\,D H\emissiontype{II} core. 
The NIR photometry of the star can not distinguish an M giant 
photosphere with the \textit{K}-band extinction $A_{K}=$~1.0~mag or 10,000~K blackbody 
with $A_{K}=$~2.4~mag \citep{1999ApJ...512..237B}. The X-ray result favors the latter 
with $N_{\rm{H}}$ compatible to the larger $A_{K}$ value. 

\subsubsection{Diffuse Source DS2}
Similarly to DS1, we first evaluate the level of contamination by point sources. 
Two bright point sources (PS2 and PS3) are found in and around DS2. 
PS3 has a different X-ray hardness than DS2,
and its contribution is minor above 2~keV (figures~\ref{fig:fitps}c and \ref{fig:fitds}b). 
However, the contamination by PS2 is significant. 
By comparing the time-averaged flux of PS2 by EPIC and DS2 by XIS, 
$\sim$ 40\% of 0.7--8.0~keV counts of DS2 is attributable to PS2.
Similar spectral shapes between PS2 and DS2 (figures~\ref{fig:fitps}b and \ref{fig:fitds}b) 
and the XIS detection of a marginal flux variation during the north field observation 
indicate that a significant fraction of the DS2 emission stems from PS2. 

\begin{figure}
  \begin{center}
    \FigureFile(80mm,60mm){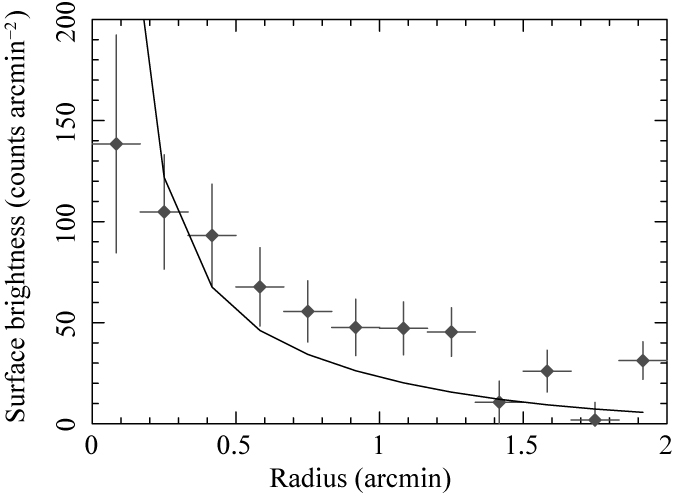}
    %%% \FigureFile(width,height){filename}
  \end{center}
  \caption{Radial profile of the background-subtracted 
intensity around PS2 (solid pluses with dots), which is 
fitted with a simulated radial profile of a point-like source at the 
same detector position (solid curve). The energy range of 2.0--5.5~keV is 
used to suppress the emission by PS3. The observed profile is constructed 
from the north field data.} \label{fig:radpro}
\end{figure}

Nevertheless, PS2 alone cannot account for all properties of DS2. DS2 has an extent
inconsistent with a single point source. Figure~\ref{fig:radpro} shows the background-subtracted 
radial profile of DS2 around PS2 in comparison to a simulated radial profile of 
a point-like source. 
The excess emission is noticeable in the profile as well as in the X-ray image (figure~\ref{fig:xisimg}c), 
where the diffuse emission is localized to the northeast of PS2.

%%%origin of excess%%%
The origin of the excess emission is uncertain. 
It can be either an ensemble of numerous unresolved point sources or the truly extended emission 
associated with the H\emissiontype{II} region. Considering a large concentration of stars 
in the Sgr\,D H\emissiontype{II} core (\cite{1999ApJ...512..237B}; \cite{2003A&A...408..127D}), 
the former possibility is likely. If it is of an extended nature, 
DS2 has quite a hard thermal spectrum for diffuse emission in H\emissiontype{II} regions, which is 
comparable to only a few cases (\cite{2002ApJ...573..191M}; \cite{2006ApJ...638..860E}). 
The metallicity (table~\ref{tab:specpar}) above one solar is larger than 
diffuse X-ray emission in other H\emissiontype{II} regions (e.g. \cite{2008PASJ...60S..85H}), 
which may be due to the enhanced metallicity in the GC region (\cite{2007ApJ...669.1011C}). 
The current data do not give a conclusive interpretation for this emission. 
Further observations are expected. 

\section{Summary}
We presented a Suzaku X-ray study of the Sgr\,D H\emissiontype{II} complex. 
XIS detected two diffuse X-ray sources and obtained their 
high signal-to-noise ratio spectra for the first time. Both 
sources have a thermal origin with emission lines 
from highly ionized atoms.

One of the diffuse sources is associated with an H\emissiontype{II} region 
and the other is a newly identified source. 
We use the X-ray characteristics to conclude that the new source is a new SNR in the 
GC region. The radio continuum data 
reinforces the idea. 

By assembling images across wavelengths, we proposed a new view of 
the Sgr\,D H\emissiontype{II} complex. It is a projection of an SNR in or 
beyond the GC and a GMC with two blister H\emissiontype{II} region 
in front of it. This revises a long-standing view that 
the Sgr\,D H\emissiontype{II} region is a single H\emissiontype{II} region 
and its environment. 

In this study, we demonstrated the capability of Suzaku XIS to investigate 
diffuse X-ray emission at low surface brightnesses. 
Along with the high-resolution radio and infrared images, the Suzaku X-ray data 
played a critical role in better understanding Sgr\,D, which is conceivably 
the case for other regions in the GC as well.

%%%%%%%%%%%%%%%%%%%%%%%%%%%%%%%%%%%%%%%
%%%acknowledgement%%%
\vspace{1pc}

The authors thank Shigeo Yamauchi at Iwate University and 
Eric D. Feigelson at Pennsylvania State University for improving the draft 
and Tomoharu Oka at Keio University for providing the CO data. 
Support for this work was provided by the National Aeronautics and Space
Administration (NASA) through Chandra Postdoctoral Fellowship Award Number
PF6-70044 issued by the Chandra X-ray Observatory Center, which is
operated by the Smithsonian Astrophysical Observatory for and on behalf
of the NASA under contract
NAS8-03060. 
Y.\,H. acknowledges financial support from the Japan Society for the Promotion of
Science. 
This work is supported by a Grants-in-Aid for the 21st Century COE 
``Center for Diversity and Universality in Physics'' 
and for Scientific Researches on Priority Areas in Japan (Fiscal Year 2002--2006) 
``New Development in Black Hole Astronomy'' and another (No.18204015)
from the Ministry of Education, Culture, Sports, Science and Technology 
of Japan. 

The research has made use of the SIMBAD database, 
operated at CDS, Strasbourg, France. 
The Digitized Sky Surveys were produced at the Space Telescope Science Institute under U.S. Government grant NAG W-2166. The images of these surveys are based on photographic data obtained using the Oschin Schmidt Telescope on Palomar Mountain and the UK Schmidt Telescope. The plates were processed into the present compressed digital form with the permission of these institutions. 
This research has also made use of the NASA/IPAC Infrared Science Archive, which is operated by the Jet Propulsion Laboratory, California Institute of Technology, under contract with the NASA.
\par

%\begin{longtable}{lll}
%  \caption{Sample of ``longtable"}\label{tab:LTsample}
%  \hline              
%  name & value1 & value2 \\ 
%\endfirsthead
%  \hline
%  name & value & value2  \\
%\endhead
%  \hline
%\endfoot
%  \hline
%\endlastfoot
%  \hline
%  aaaaa & bbbbb & ccccc \\
%  ...... & ..... & ..... \\
%  ...... & ..... & ..... \\
%  ...... & ..... & ..... \\
%  xxxxx & yyyyy & zzzzz \\
%\end{longtable}
%
%\appendix
%\section{Method of .....}
%
%\section{Approximation of ...}
%
%\section*{Complete data}

%%%
% See the manual for the detail.
%%%

\end{document}